\documentclass[11pt,a4paper]{article}

\usepackage{graphicx}
\usepackage{epsfig}

\usepackage{hyperref}

\usepackage{color}

\usepackage{graphicx,epsf}
\usepackage{bm}
\usepackage{amsmath}

\usepackage[dvipsnames]{xcolor}

\def \d {\mathrm{d}}
\begin{document}

\begin{center}
{\large{\textbf{Saturation of energetic-particle-driven geodesic acoustic modes due to wave-particle nonlinearity}}}\\
\vspace{0.2 cm}
{\normalsize {\underline{A. Biancalani}$^1$, I. Chavdarovski$^1$, Z. Qiu$^2$, A. Bottino$^1$, D. Del Sarto$^3$, A. Ghizzo$^3$, \"O G\"urcan$^4$, P. Morel$^4$, I. Novikau$^1$.\\}}
\vspace{0.1 cm}
\small{\it
$^1$ Max-Planck Institute for Plasma Physics, 85748 Garching, Germany\\
$^2$ Institute for Fusion Theory and Simulation and Department of Physics, Zhejiang University, 310027 Hangzhou, People's Republic of China\\
$^3$ Institut Jean Lamour- UMR 7198, University of Lorraine, BP 239 F-54506 Vandoeuvre les Nancy, France\\ 
$^4$ LPP, CNRS, ​\`Ecole polytechnique, UPMC Univ Paris 06, Univ. Paris-Sud, Observatoire de Paris, Universit\'e Paris-Saclay, Sorbonne Universit\'es, PSL Research University, 91128 Palaiseau, France\\
}
\small{
contact of main author: \url{www2.ipp.mpg.de/~biancala}\\
}
\end{center}

\begin{abstract}
The nonlinear dynamics of energetic-particle (EP) driven geodesic acoustic modes (EGAM) is investigated here. A numerical analysis with the global gyrokinetic particle-in-cell code ORB5 is performed, and the results are interpreted with the analytical theory, in close comparison with the theory of the beam-plasma instability. Only axisymmetric modes are considered, with a nonlinear dynamics determined by wave-particle interaction.
Quadratic scalings of the saturated electric field with respect to the linear growth rate are found for the case of interest. The EP bounce frequency is calculated as a function of the EGAM frequency, and shown not to depend on the value of the bulk temperature.
Near the saturation, we observe a transition from adiabatic to non-adiabatic dynamics, i.e., the frequency chirping rate becomes comparable to the resonant EP  bounce frequency. The numerical analysis is performed here with electrostatic simulations with circular flux surfaces, and kinetic effects of the electrons are neglected.
\end{abstract}

\section{Introduction}

Two main issues related to magnetic confinement fusion are the turbulent transport, and the dynamics of energetic particles (EP), produced by fusion reactions, or injected for heating purposes. Zonal (i.e. axisymmetric) electric fields are observed to interact with turbulence in tokamaks, in the form of zero-frequency zonal flows (ZF)~\cite{Hasegawa79,Rosenbluth98,Diamond05} and finite-frequency geodesic acoustic modes (GAM)~\cite{Winsor68,Zonca08}. Geodesic acoustic modes, due to their finite frequency, can also interact with EP via inverse Landau damping, leading to EP-driven GAM (EGAM) given that EP drive is sufficient to overcome the threshold condition induced by GAM Landau damping and continuum damping~\cite{Fu08,Qiu10,Qiu11,Qiu12,Zarzoso12,Wang13,Girardo14}.
Understanding the dynamics of EGAMs is crucial due to their interaction with turbulence, which can modify the turbulent transport~\cite{Zarzoso13,Zarzoso17}. Moreover, a strong nonlinear interaction of EGAM with EP is observed in tokamaks~\cite{Lauber14,Horvath16}, leading potentially to a strong redistribution of the EP in phase space.

In order to predict the importance of the interaction of EGAMs with turbulence and with EP, it is important to understand their nonlinear saturation mechanisms. One of the reasons of saturation for EGAMs is the wave-particle nonlinear interaction. In this case, the saturation occurs due to the EP nonlinear redistribution in phase space, and the consequent decrease of the energy exchange between the EP and the EGAM. Another possible reason of saturation is the wave-wave coupling. This can occur for an EGAM interacting with another EGAM and generating zonal side-bands, or for an EGAM with non-zonal modes, e.g., turbulence~\cite{Zarzoso13}. In this paper, we focus on the wave-particle nonlinear interactions. For an investigation of the EGAM self-coupling, see Ref.~\cite{Qiu17}.

The nonlinear saturation of EGAMs is investigated here by means of electrostatic simulations with the gyrokinetic particle-in-cell code ORB5~\cite{Jolliet07,Bottino11,Bottino15JPP}.  ORB5 has been succesfully verified against analytical theory and benchmarked against other gyrokinetic codes, for the linear dynamics of GAMs~\cite{Biancalani17veri}, and EGAMs~\cite{Biancalani14,Zarzoso14}. A detailed comparison with the beam-plasma instability (BPI)~\cite{Oneil65,Oneil68} in a 1-dimensional uniform plasma is also done, following the scheme anticipated in Ref.~\cite{Qiu11,Qiu14}.
Similarly to the BPI, the saturation level of EGAM is shown to scale quadratically with the linear growth rate.
A similar investigation was also previously done for Alfv\'en modes (see, for example, Ref.~\cite{Briguglio14}). The EGAM frequency is shown to evolve in time when approaching the saturation, like the BPI (see, for example, Ref.~\cite{Morales72,Armon16}). The chirping rate is observed to get of the order of magnitude of the squared of the EP bounce frequency near the saturation. This denotes a transition from adiabatic to non-adiabatic regimes.

The paper is organized as follows. The adopted gyrokinetic model is described in Sec.~\ref{sec:model}, and the equilibrium and case definition in Sec.~\ref{sec:equil}. The linear dynamics is described in Sec.~\ref{sec:linear}. The saturation levels are investigated in Sec.~\ref{sec:scalings}. The regimes of different adiabaticity are investigated by means of the analysis of the frequency, in Sec.~\ref{sec:frequency}.  Finally, a summary of the results is given in Sec.~\ref{sec:conclusions}.

\section{The model}
\label{sec:model}

The main damping mechanism of GAM and EGAM is the Landau damping, which makes the use of a kinetic model necessary. In this paper we use the global gyrokinetic particle-in-cell code ORB5~\cite{Jolliet07}. ORB5 was originally developed for electrostatic ITG turbulence studies, and recently extended to its multi-species electromagnetic version in the framework of the NEMORB project~\cite{Bottino11,Bottino15JPP}. In this paper, collisionless electrostatic simulations are considered.

The model equations of the electrostatic version of ORB5 is made by the trajectories of the markers, and by the gyrokinetic Poisson law for the scalar potential $\phi$. These equations are derived in a Lagrangian formulation~\cite{Bottino15JPP}, based on the gyrokinetic Vlasov-Maxwell equations of Sugama, Brizard and Hahm~\cite{Sugama00,Brizard07}.
The equations for the marker trajectories (in the electrostatic version of the code) are~\cite{Bottino15JPP}:
\begin{eqnarray}
\dot{\bf R}&=&\frac{1}{m_s}p_\|\frac{\bf{B^*}}{B^*_\parallel} + \frac{c}{q_s B^*_\parallel} {\bf{b}}\times \left[\mu \nabla B + q_s \nabla  \tilde\phi  \right]  \label{eq:traj-1}\\
\dot{p_\|}&=&-\frac{\bf{B^*}}{B^*_\parallel}\cdot\left[\mu \nabla B + q_s
  \nabla  \tilde\phi  \right] \label{eq:traj-2}\\
 \dot{\mu} & = & 0 \label{eq:traj-3}
\end{eqnarray}
The coordinates used for the phase space are $({\bf{R}},p_\|,\mu)$, i.e. respectively the gyrocenter position, canonical parallel momentum $p_\| = m_s v_\|$ and magnetic momentum $\mu = m_s v_\perp^2 / (2B)$ (with $m_s$ and $q_s$ being the mass and charge of the species).  $v_\|$ and $v_\perp$ are respectively the parallel and perpendicular component of the particle velocity.
The gyroaverage operator is labeled here by the tilde symbol $\tilde{}$. ${\bf{B}}^*= {\bf{B}} + (c/q_s)  {\bf{\nabla}}\times ({\bf{b}} \, p_\|)$, where ${\bf{B}}$ and ${\bf{b}}$ are the equilibrium magnetic field and magnetic unitary vector.

Kinetic effects of the electrons are neglected. This is done by calculating the electron gyrocenter density directly from the value of the scalar potential as~\cite{Bottino15JPP}:
\begin{equation}
n_e({\bf{R}},t) = n_{e0} + \frac{q_s n_{e0}}{T_{e0}} \big( \phi - \bar\phi \big)   \label{eq:adiabatic-electrons}
\end{equation}
where $\bar\phi$ is the flux-surface averaged potential, instead of treating the electrons with markers evolved with Eqs.~\ref{eq:traj-1}, \ref{eq:traj-2}, \ref{eq:traj-3}.

We focus on the dynamics of zonal perturbations, by filtering out all non-zonal components (this is to avoid interactions of zonal/non-zonal modes). Wave-wave coupling is neglected, by evolving the bulk-ion and electron markers along unperturbed trajectories. This means that, in Eqs.~\ref{eq:traj-1}, \ref{eq:traj-2}, \ref{eq:traj-3} for the bulk ions, the last terms, proportional to the EGAM electric field, are dropped. The nonlinear wave-particle dynamics is studied by evolving the EP markers along the trajectories which include perturbed terms associated with the EGAM electric field. This means that the EP markers are evolved with  Eqs.~\ref{eq:traj-1}, \ref{eq:traj-2}, \ref{eq:traj-3} where the terms proportional to the EGAM electric field are retained.

The gyrokinetic Poisson equation is~\cite{Bottino15JPP}:
\begin{equation}
 - {\bf{\nabla}} \cdot \frac{n_0 m_i c^2}{B^2} \nabla_\perp \phi=  \sum_{i} \int \d W  q_s \, \tilde{\delta f_s}  -n_e({\bf{R}},t)\label{eq:Poisson}
\end{equation}
with $n_0 m_i$ being here the total plasma mass density (approximated as the ion mass density). The summation over the species is performed for the bulk ions and for the EP, whereas  the electron contribution is given by $-n_e({\bf{R}},t)$.
Here $\delta f = f - f_0$ is the gyrocenter perturbed distribution function, with $f$ and $f_0$ being the total and equilibrium (i.e. independent of time, assumed here to be a Maxwellian) gyrocenter distribution functions.
The integrals are over the phase space volume, with $\d W =(2\pi/m_i^2) B_\|^* \d p_\| \d \mu$ being the velocity-space infinitesimal volume element.

\section{Equilibrium and simulation parameters}
\label{sec:equil}

The tokamak magnetic equilibrium is defined by a major and minor radii of $R_0=1$ m and $a=0.3125$ m, a magnetic field on axis of $B_0=1.9$ T, a flat safety factor radial profile, with $q=2$, and circular flux surfaces (with no Grad-Shafranov shift).
Flat temperature and density profiles are considered at the equilibrium. The bulk plasma temperature is defined by $\rho^*=\rho_s/a$, with $\rho_s = c_s/\Omega_i$, with $c_s = \sqrt{T_e/m_i}$ being the sound speed and $\Omega_i$ the cyclotron frequency. Three increasing values of bulk plasma temperature are considered, for investigating the dependence of our results on the Landau damping, corresponding to: $\rho^*_1= 1/256=0.0039$, $\rho^*_2= 1/128=0.0078$, and $\rho^*_3= 1/64=0.0156$ ($\tau_e=T_e/T_i=1$ for all cases described in this paper).

All parameters defined so far, are adopted by ORB5 with no consideration of the mass of the bulk ion species. The choice of the bulk ion mass is done only during the post-processing of the results of ORB5, if we need to know the values of equilibrium or perturbed quantities in non-normalized units.
In particular, in the case of a hydrogen plasma, we get a value of the ion cyclotron frequency of $\Omega_i = 1.82 \cdot 10^8 rad/s$. With this choice of bulk specie, we can calculate the plasma temperature and sound frequency.
The three values of plasma temperature are $T_{i1}=$ 515 eV, $T_{i2}=$ 2060 eV, and $T_{i3}=$ 8240 eV. The sound frequency is defined as $\omega_s = 2^{1/2} v_{ti}/R$ (with $v_{ti} = \sqrt{T_i/m_i}$, which for $\tau_e=1$ reads $v_{ti}=c_s$). We obtain the following three values of the sound velocity: $c_{s1} = 2.22\cdot 10^5$ m/s, $c_{s2} = 4.44 \cdot 10^5$ m/s, $c_{s1} = 8.88\cdot 10^5$ m/s. These correspond to the following three values of the sound frequency: $\omega_{s1} = 3.14 \cdot 10^5$ rad/s, $\omega_{s2} = 6.28 \cdot 10^5$ rad/s, and $\omega_{s3} = 1.25 \cdot 10^6$ rad/s.

\begin{figure}[t!]
\begin{center}
\includegraphics[width=0.51\textwidth]{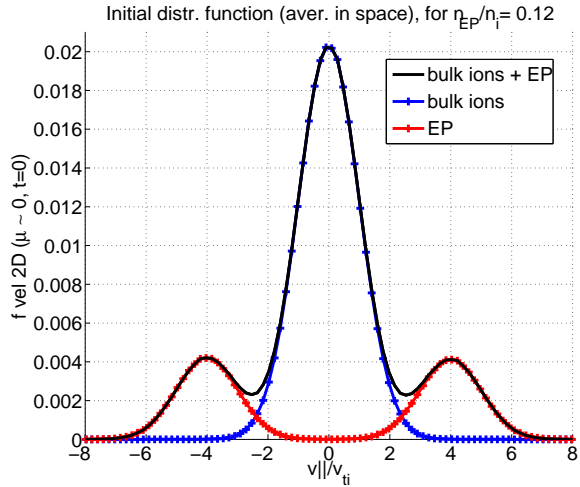}
\caption{Initial EP distribution function for a simulation with $n_{EP}/n_i = 0.12$, $v_{bump}/v_{ti}$ = 4.\label{fig:fvel2D_t0}}
\end{center}
\end{figure}

The energetic particle distribution function is a double bump-on-tail, with two bumps at $v_\| = \pm v_{bump}$ (see Fig.~\ref{fig:fvel2D_t0}), like in Ref.~\cite{Biancalani14}. In this paper, $v_{bump}=4 \, v_{ti}$ is chosen.
In order to initialize a distribution function which is function of constants of motion only, the modified variable
$\tilde{v}_\| = \sqrt{2(E-\mu B_{max})/m/v_{ti}}$ is used instead of $v_\|= \sqrt{2(E-\mu B(r,\theta))/m/v_{ti}}$ (similarly to Ref.~\cite{Zarzoso12,Biancalani14,Zarzoso14}).
Neumann and Dirichlet boundary conditions are imposed to the scalar potential, respectively at the inner and outer boundaries, $s=0$ and $s=1$.

\section{Linear dynamics}
\label{sec:linear}

\begin{figure}[t!]
\begin{center}
\includegraphics[width=0.49\textwidth]{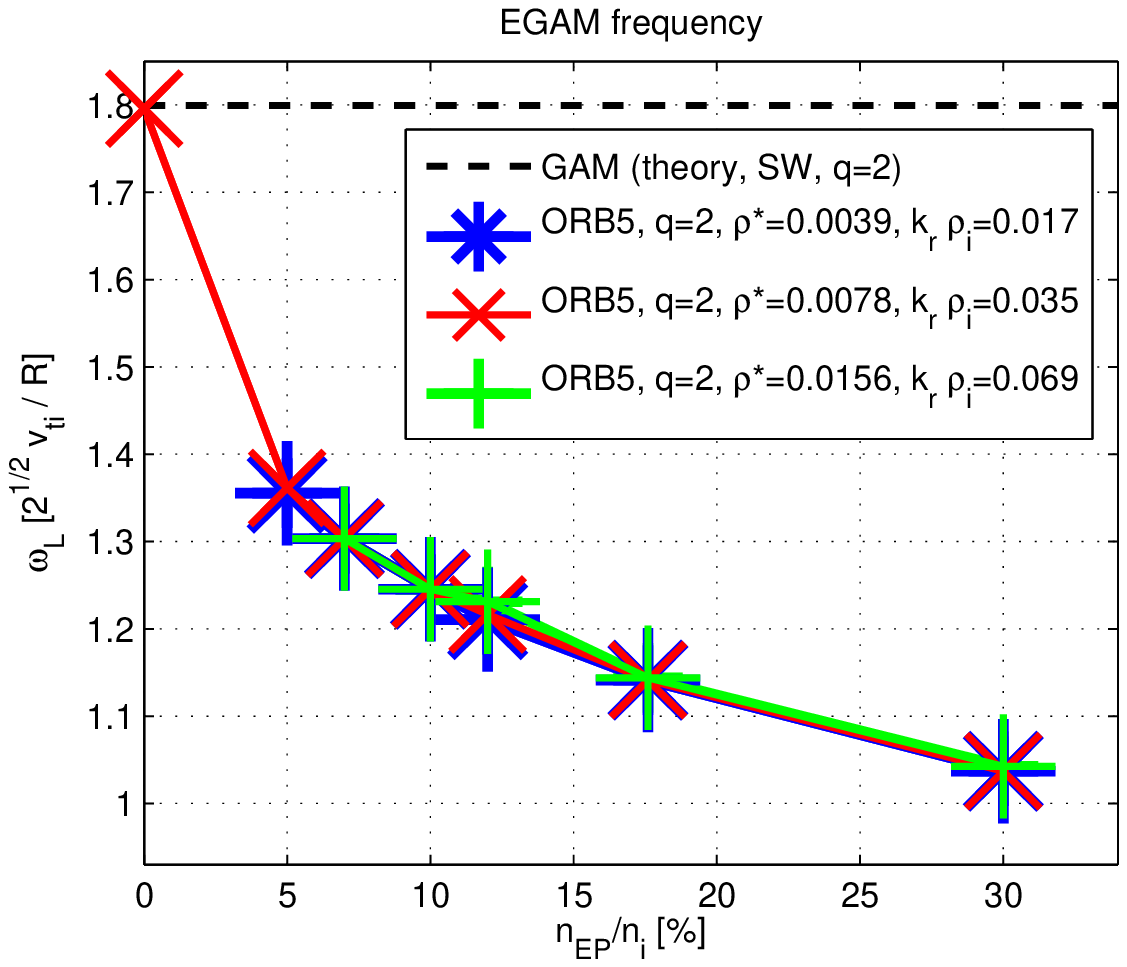}
\includegraphics[width=0.49\textwidth]{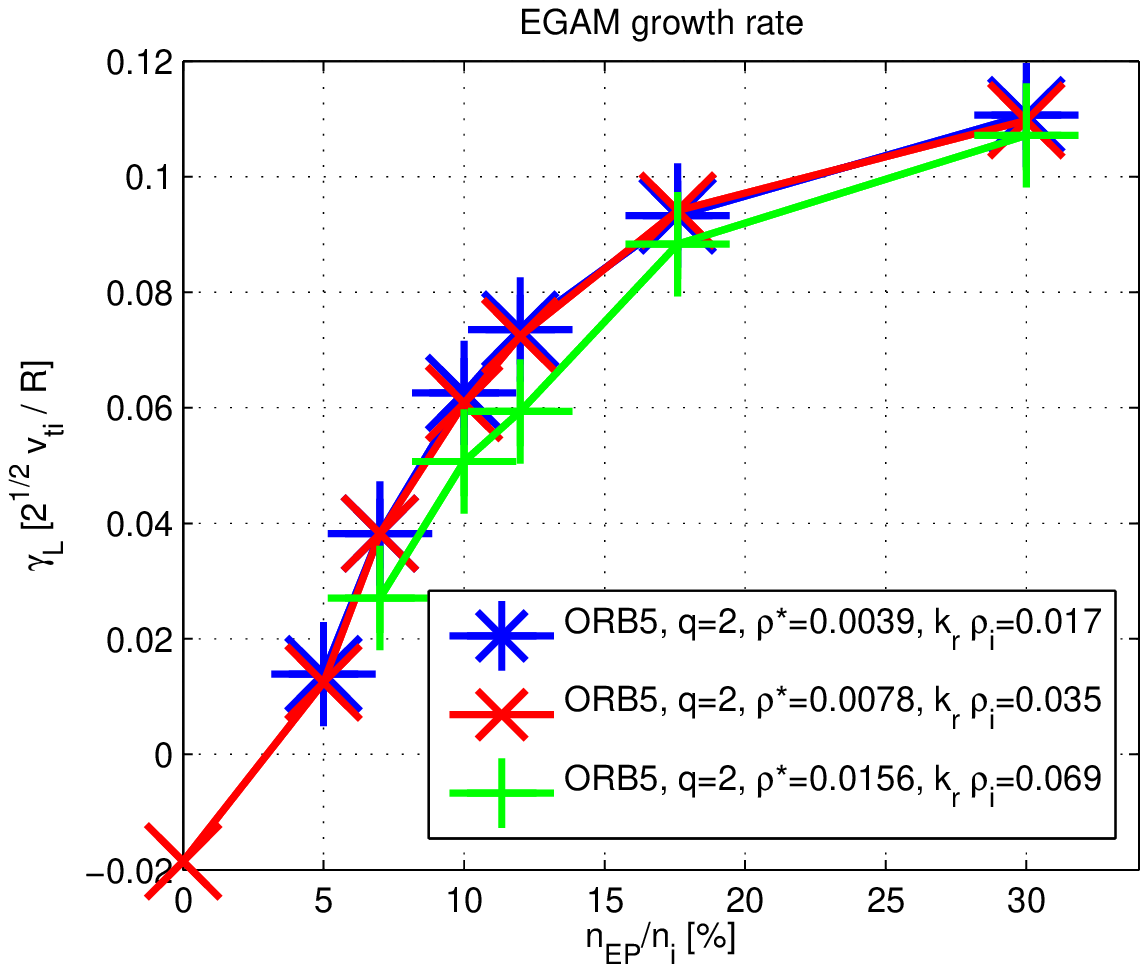}
\caption{Frequency (left) and growth rate (right) vs EP concentration, for simulations with $\bar\zeta=v_{bump}/v_{th}$=4.\label{fig:omegagamma_nEP}}
\end{center}
\end{figure}

The linear dynamics of EGAMs in an equilibrium similar to the one considered in Sec.~\ref{sec:equil}, has been investigated in Ref.~\cite{Zarzoso14} for a bulk plasma temperature given by $\rho^*=1/64=0.0156$, by means of ORB5 simulations and analytical theory.
A scan on the EP concentration is performed here, similarly to what was done in Ref.~\cite{Zarzoso14}, but with three different values of bulk plasma temperature, corresponding to three different values of $\rho^*$, as described in Sec.~\ref{sec:equil}.
The dependence of the linear dynamics (frequency and growth rate) on the EP concentration is shown in Fig.~\ref{fig:omegagamma_nEP}.
Both the frequency and the growth rates are observed to follow the qualitative scalings as described in Ref.~\cite{Biancalani14} and \cite{Zarzoso14}. Note, in particular, that the growth rate does not grow linearly with $n_{EP}$.

The dependence of the frequency on the EP concentration is not observed to change with $\rho^*$. Regarding the growth rate, no change is observed when going from $\rho^*_3 = 0.0039$ to $\rho^*_2 = 0.0078$, meaning that the measured growth rate is basically given by the value of the drive, and the Landau damping here is negligible for the chosen values of EP concentration. On the other hand, when further increasing the temperature, and going to  $\rho^*_1 = 0.0156$, a smaller value of growth rate is measured, meaning a higher Landau damping.
The transit resonance velocity of the EP can be calculated by knowing the EGAM frequency of a specific simulation. Considering a case with $n_{EP}/n_i = 0.12$ as an example, the frequency is measured as: $\omega_L= 1.2 \, \omega_s$. For comparison, the GAM frequency for these parameters is $\omega_{GAM}= 1.8 \, \omega_s$. Then, the transit resonance velocity in the linear phase is calculated as $v_{\|0} = qR \, \omega_L = 3.4 \, v_{ti}$, with $\omega_L$ being the EGAM linear frequency.

\section{Scaling of the saturated amplitudes}
\label{sec:scalings}

In this Section, we focus on the value of the saturated electric field $\bar{\delta{E_r}}$, and we investigate its dependence on the value of the linear growth rate and of the damping.
The corresponding scaling of $\bar{\delta{E_r}}$ sheds light on the mechanism which is responsible for the saturation. A  one-to-one comparison with the saturation mechanism of the BPI is also described.

\begin{figure}[b!]
\begin{center}
\includegraphics[width=0.48\textwidth]{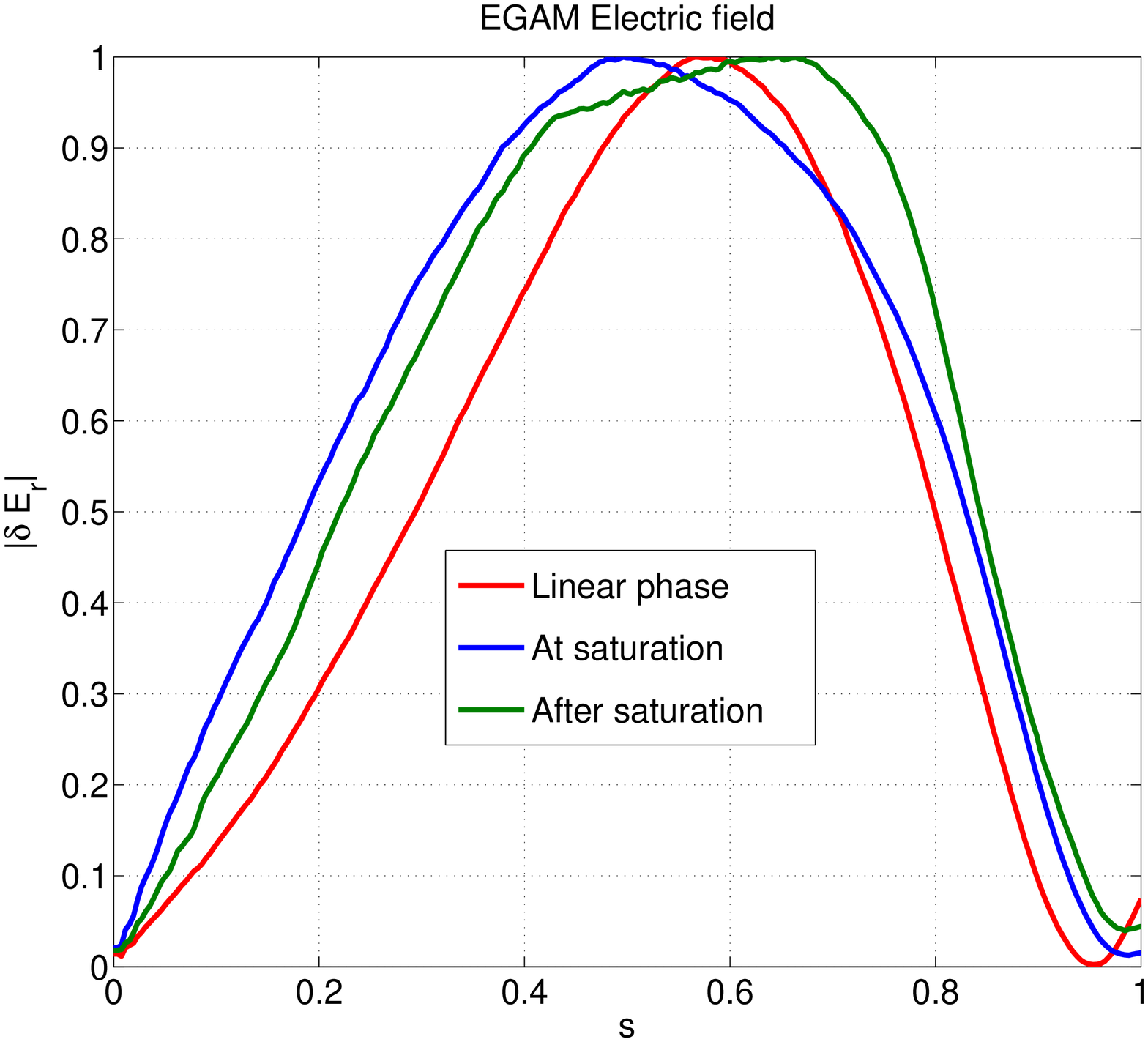}
\includegraphics[width=0.50\textwidth]{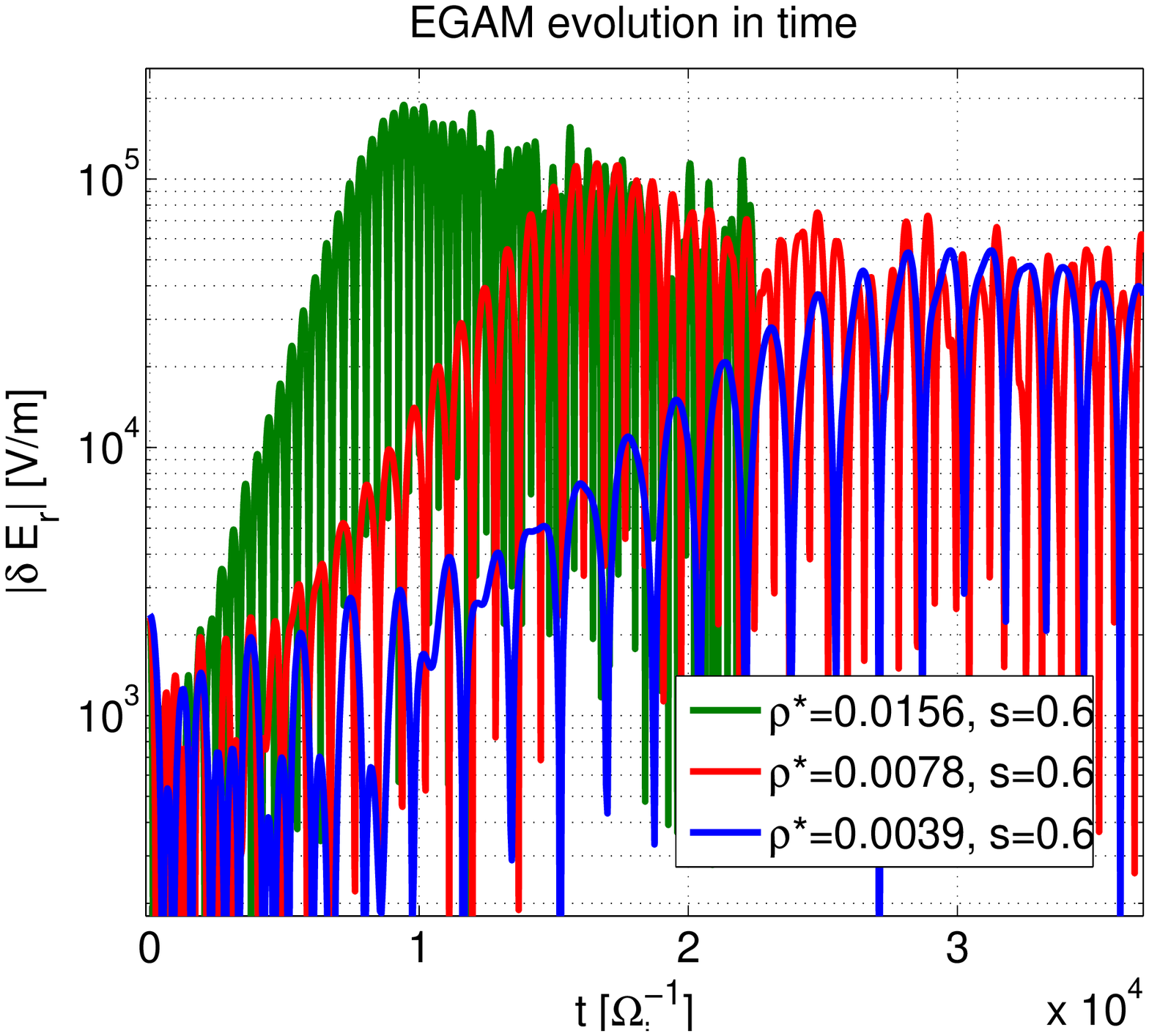}
\caption{On the left, EGAM normalized radial structure for $\rho^*=0.0156$, $n_{EP}/n_i=0.176$. On the right, absolute value of the electric field measured at the position of the peak, $s=0.6$, for three different simulations with respectively $\rho^*=0.0039$ (blue), $\rho^*=0.0078$ (red), $\rho^*=0.0156$ (green). All simulations here have $n_{EP}/n_i=0.30$. The time is expressed in units of $\Omega_i^{-1}$.\label{fig:nonlinear-Omi}}
\end{center}
\end{figure}

The amplitude of the EGAM at saturation has been measured in different simulations performed with ORB5, for different values of the bulk temperature, and different values of the EP concentrations. As an example, the radial structure of a nonlinear simulation with $\rho^*=0.0156$, $n_{EP}/n_i=0.176$ is depicted in Fig.~\ref{fig:nonlinear-Omi}-a. No sensible change in the radial wave-number is observed when going from the linear phase, to the saturation, and after the saturation. This confirms that in this particular configuration, where all equilibrium radial profiles are flat, EGAM can be treated as a 1-dimensional problem where the radial direction does not play an important role.

When varying only the bulk temperature, both the linear frequency and growth rate are observed to scale with the sound frequency, which is a good normalization frequency (consistently with Fig.~\ref{fig:omegagamma_nEP}). The saturation level increases with the linear growth rate, similarly to other instabilities like the BPI in a uniform system and the Alfv\'en instabilities in tokamaks. This is depicted in Fig.~\ref{fig:nonlinear-Omi}-b, where non-normalized units are used (in particular, the ion cyclotron frequency is selected as a time unit not depending on the temperature).

\begin{figure}[t!]
\begin{center}
\includegraphics[width=0.485\textwidth]{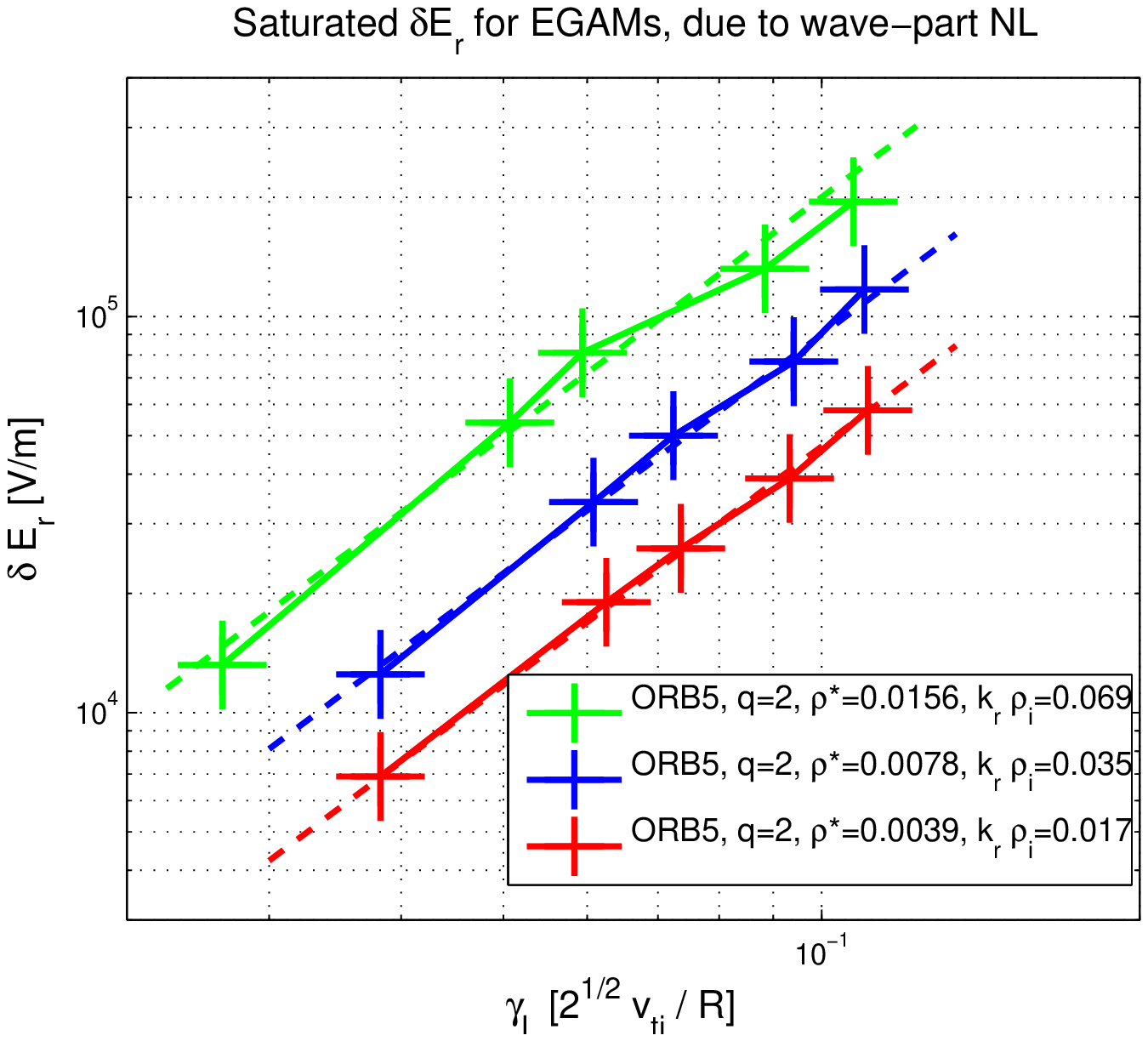}
\includegraphics[width=0.49\textwidth]{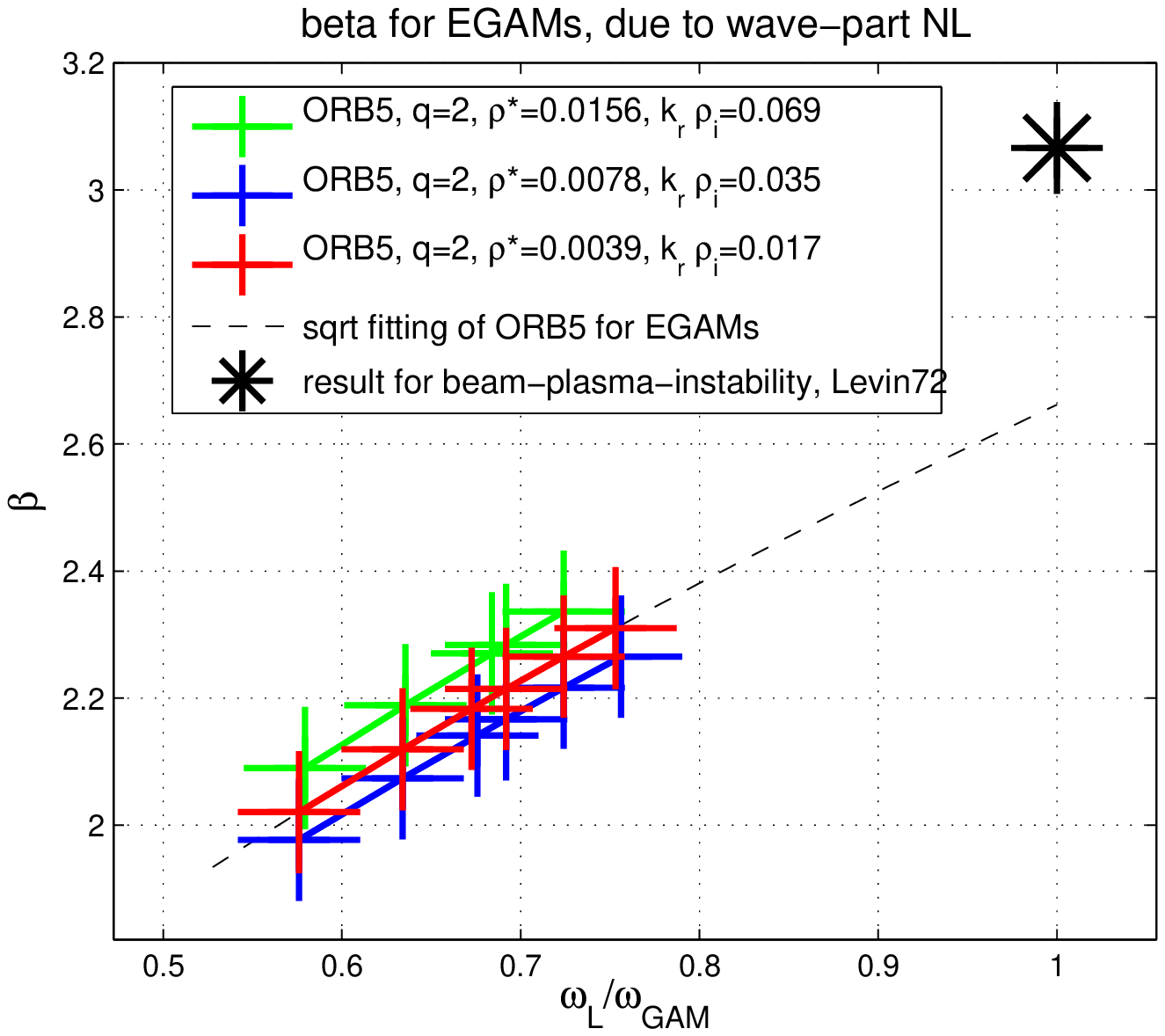}
\caption{On the left, maximum value of the EGAM radial electric field, vs linear growth rate, for the same simulations as in Fig.~\ref{fig:omegagamma_nEP}. The red, blue and green crosses refer respectively to $\rho^*=0.0039$, $\rho^*=0.0078$, $\rho^*=0.0156$. The dashed lines are the quadratic fitting formulas. On the right, the value of $\beta$ as given in Eq.~\ref{eq:beta}, vs the linear frequency, for the same simulations. The black dashed line is the sqruare root interpolation. For a reference, the black star shows the result obtained for the BPI in Ref.~\cite{Levin72}. \label{fig:E_gamma}}
\end{center}
\end{figure}

The scalings with the energetic particle concentration are also investigated. The results are shown in Fig.~\ref{fig:E_gamma}a. We obtain that, in the considered regime, the saturated level scales as the quadratic power of the linear growth rate.
This quadratic scaling is typical for marginally stable bump-on-tail instabilities, as derived by O'Neil~\cite{Oneil65,Oneil68}.

We can consider the problem to be similar to a monochromatic beam-plasma system, in which particles are moving in the potential well of the perturbed electric field. Depending on the energy, some particles are trapped inside the well and execute bounce motion with frequency $\omega_b$ in the frame moving with the wave phase velocity.
The resonant particles exchange energy with the mode, causing the amplitude to grow and the particles to redistribute in phase space, flattening the velocity distribution in the vicinity of the resonant parallel velocity $v_\parallel= \omega q R_0$.
The drive is due to the positive slope of the particle distribution in the velocity space at the resonant parallel velocity, which acts as an inverse Landau damping.
In the initial stage $\omega_b \ll \gamma_L$, the mode grows exponentially with a linear growth rate, making  more and more particles to become trapped in the phase space.
After some significant particle velocity redistribution, the power exchange between the wave and the particles is balanced, causing the wave amplitude to saturate.

Since the initial perturbation is negligible, the saturation level is determined by the exchange of energy between the mode and a band of resonant particles~\cite{Levin72}.
The chirping of the mode seen in Fig.~\ref{fig:NL-freq}  is a strongly non-linear effect that occurs when the amplitude is large enough to have the trapped (and more generally resonant) particles with $\omega_b \sim \gamma_L$ drastically change the dynamics of the mode through the modification of the distribution function  and non-perturbative fast particle response.
Namely, the mode dynamics is determined by all resonant particles that exhibit a continuous oscillation, trapping and detrapping in the potential well of the mode, thus contributing (non-perturbatively) to the non-adiabatic behavior observed in the simulation.

The quadratic scaling of the saturation level with the damping obtained with our simulations  is similar to the saturation of the BPI, where it occurs due to wave-particle trapping~\cite{Oneil65}. For the BPI, the original reference of M. B. Levin (1972) gives a value of $\omega_b = 3.06 \, \gamma_L$ at saturation~\cite{Levin72} (for comparison, note that more recent numerical calculations find $\omega_b = 3.2 \, \gamma_L$~\cite{Lesur09,Carlevaro17}). For EGAMs, the bounce frequency is given by~\cite{Qiu11}:
\begin{equation}\label{eq:omegab_vs_E}
\omega_b^2 = \alpha_1 \, \bar{\delta{E_r}} \, , \;\;\;\; \text{with} \; \alpha_1 \equiv \frac{ e \hat{V}_{dc}}{2 m_{EP} v_{\|0} q R_0}
\end{equation}
with $m_{EP}$ being the mass of the energetic particle specie, considered equal to the bulk ion mass in this paper, $v_{\|0}$ the velocity matching the resonance condition, and $\hat{V}_{dc} = m_{EP} v^2_{\|0}/(eB R)$ the magnetic curvature drift. Therefore we have:
\begin{equation}\label{eq:alpha_1}
\alpha_1 = \frac{v_{\|0}}{2 q R^2 B}  = \frac{\omega_L}{2 R B}
\end{equation}
We emphasize that the value of $\alpha_1$ depends on  $\omega_L$. This is a main difference with respect to the BPI, where there is only one value $\omega_{lin}=\omega_{pe}$, with $\omega_{pe}$ being the plasma frequency.

The dependence of the maximum electric field on the linear growth rate can be measured with the results of the numerical simulations. For the simulations shown in Fig.~\ref{fig:E_gamma}, we find:
\begin{equation}
\bar{\delta{E_r}} = \alpha_2 \gamma_L^2
\end{equation}
The values of $\alpha_2$ are found to depend on the bulk temperature. For the three chosen increasing values of $\rho^*$, i.e. $\rho^* =$ 0.0039, 0.0078 and 0.0156, we have respectively $\alpha_2= 0.47\cdot 10^7$ V/m, $0.9\cdot 10^7$ V/m, and $2.0\cdot 10^7$ V/m. Finally the relationship between the EP bounce frequency and the linear growth rate is obtained:
\begin{equation}
 \omega_b = \beta \, \gamma_L
 \label{eq:omegab_vs_gammalin}
\end{equation}
where $\beta$ is calculated as $\beta = (\alpha_1 \alpha_2 )^{1/2}/\omega_s$, which yields:
\begin{equation}
\beta = \beta_0 \Big(\frac{\omega_L}{\omega_{GAM}}\Big)^{1/2}, \;\; \text{with} \;\; \beta_0 = \frac{1}{\omega_s}\Big( \frac{\omega_{GAM} \, \alpha_2}{2RB} \Big)^{1/2} \label{eq:beta}
\end{equation}
Note that here, $\beta$ depends on the EGAM frequency, which changes with the intensity of the drive, given here by the EP concentration (see Fig.~\ref{fig:omegagamma_nEP}). As a comparison, note that in the problem of the BPI, solved in Ref.~\cite{Oneil65,Oneil68,Levin72}, on the other hand, the mode frequency is assumed to be constant and equal to the frequency measured in the absence of EP.
On the other hand, the values of $\beta$ are found not to depend on the damping, which changes with the three different bulk temperatures: an interpolation can be drawn for all considered simulations, and shown to depend on $\omega_L$ only (see Fig.~\ref{fig:E_gamma}b). In this sense, the formula given in Eq.~\ref{eq:beta} is universal for the chosen regime, because it does not depend on the bulk plasma temperature.

The considered equilibrium has been chosen in order to excite EGAMs out of a GAM, and not out of a Landau pole, as described in Ref.~\cite{Zarzoso14}. This choice of the mode is done, in order to have a one-to-one correspondence with the BPI, where the mode which is excited by the energetic electron beam is the Langmuir wave which is an eigenmode of the system in the absence of EP.
Following this consideration, we can consider the interpolation of the results shown in Fig.~\ref{fig:E_gamma}, and take the extrapolation to $\omega_L\rightarrow \omega_{GAM}$, which is the limit assumed in the resolution of the BPI. In this case, the extrapolation gives a unique value, which defines the EGAM instability, i.e. $\beta_0 = \beta(\omega_L/ \omega_{GAM} =1) = 2.66$.  This is to be compared with the value of $\beta$ obtained for the BPI~\cite{Levin72,Lesur09,Carlevaro17}, i.e. $\beta_{BPI}=3.2$ (originally estimated as $\beta_{BPI}=3.06$ by Levin).

Finally, by using Eqs.~\ref{eq:omegab_vs_E}, \ref{eq:alpha_1}, \ref{eq:omegab_vs_gammalin} and \ref{eq:beta}, we can write the formula for the saturated electric field as a function of the linear characteristics of the mode:
\begin{equation}
\delta E_r = \frac{2 R B \beta_0^2}{\omega_{GAM}} \; \gamma_L^2
\end{equation}
with the value of $\beta_0 = 2.66$ in the regime considered in this paper.

\section{Frequency}
\label{sec:frequency}

In this section, we show the results of the measurement of the time evolution of the EGAM frequency. In Sec.~\ref{sec:scalings}, we have shown that a quadratic scaling of the saturated electric field on the linear growth rate is found. This quadratic scaling, has a one-to-one correspondence on the Langmuir wave problem investigated by O'Neil, where the saturation occurs due to wave-particle trapping. The wave-particle trapping mechanism, is usually referred to as adiabatic, meaning that a slowly increasing potential well traps more and more energetic particles. In this adiabatic regime, the mode frequency varies very slowly with respect to the bounce frequency. On the other hand, in the EGAM case considered here, we show that the saturation is not strictly adiabatic, but a transition between adiabatic and non-adiabatic regime occurs at the time of the saturation.

\begin{figure}[t!]
\begin{center}
\includegraphics[width=0.51\textwidth]{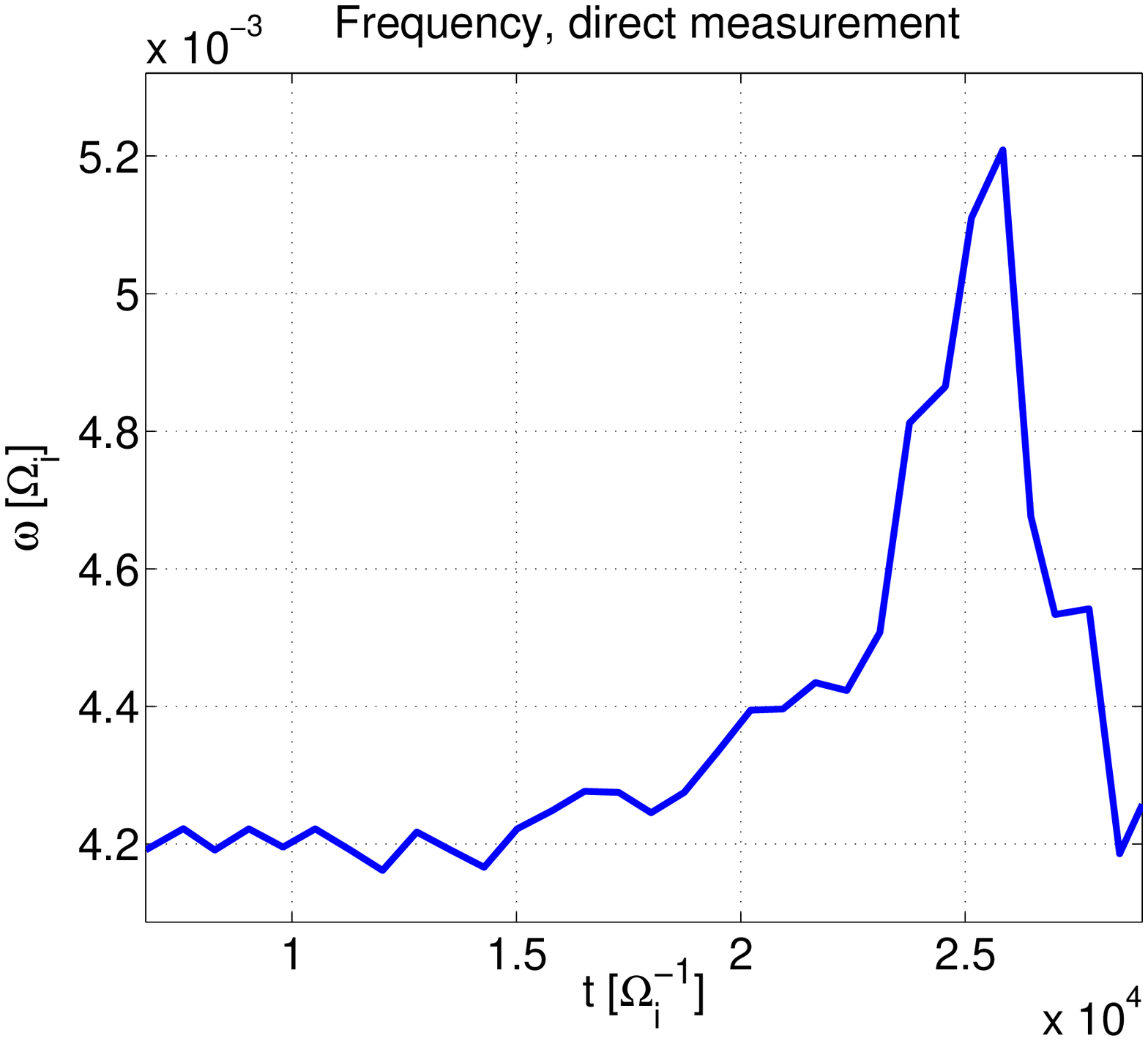}
\includegraphics[width=0.47\textwidth]{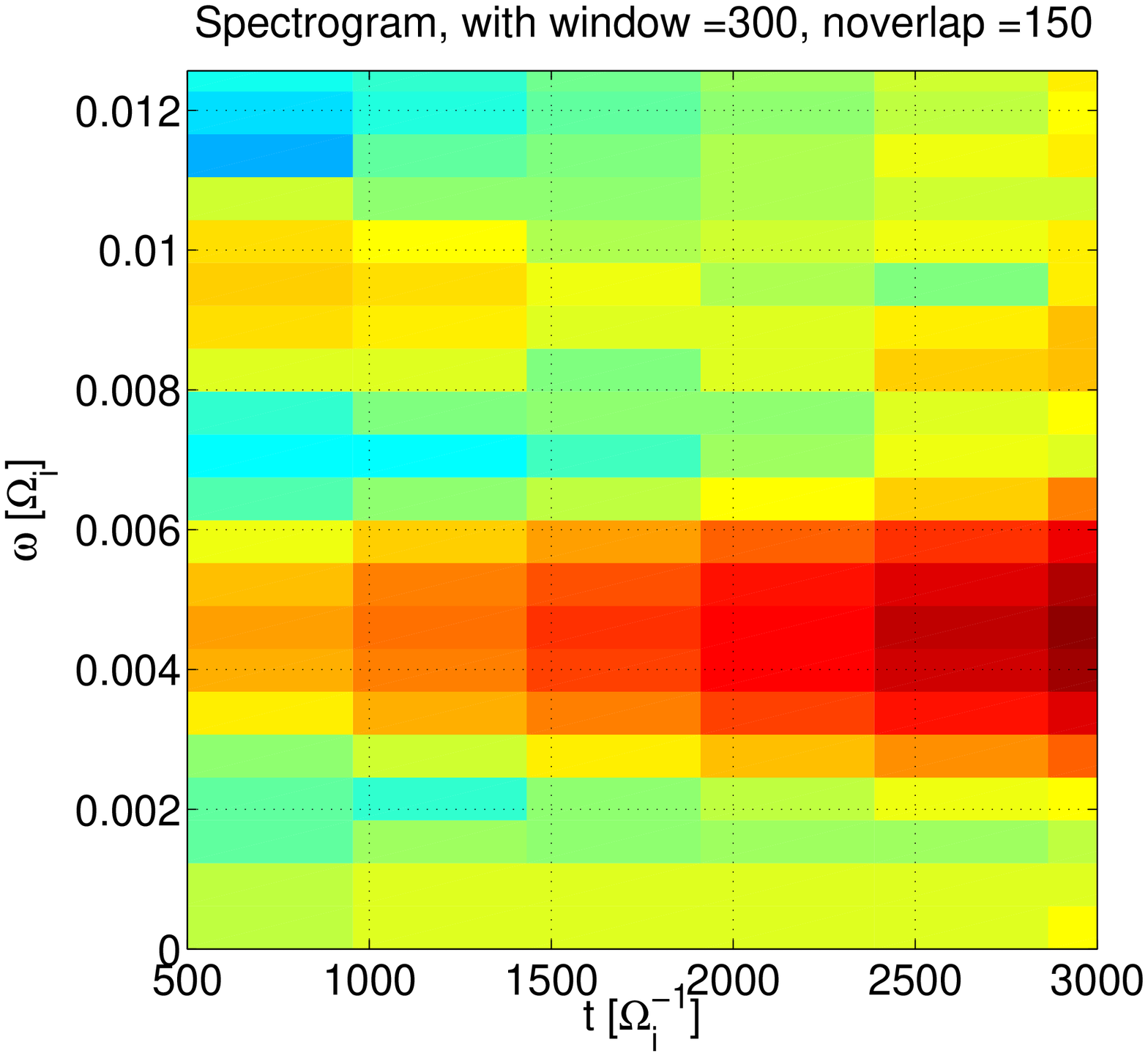}
\caption{Nonlinear evolution of the frequency, measured as a short-time average of the period between the peaks (left) or with a short-time Fourier transform (right), for $\rho^*=0.0078$, $n_{EP}/n_i=0.12$.\label{fig:NL-freq}}
\end{center}
\end{figure}

In this section, we take again the EGAM case with $\rho^*=0.0078$, $n_{EP}/n_i=0.12$, as a typical case, and we investigate the variation of the frequency in time in comparison with the bounce frequency. For the measurement of the frequency, we use the radial zonal electric field  measured at s=0.5. The measurement of the frequency is performed in two different ways: a) as an average of the period between several EGAM oscillation peaks, as shown in Fig.~\ref{fig:NL-freq}-a; b) with a short-time Fourier transform (STFT), as shown in Fig.~\ref{fig:NL-freq}-b.

With the first technique, namely measuring the frequency by inversion of the period between neighbouring peaks, an upward chirping is observed in the nonlinear phase, of the order of 10\% of the linear frequency. This means that the resonance condition changes in time, with resonance velocity slightly increasing at the time of the saturation or in the later phase (see Fig.~\ref{fig:NL-freq}-a). A dominantly upward chirping of EGAMs was previously observed and documented in Ref.~\cite{Berk06,Wang13}.

The second technique, consists in measuring the frequency with a short-time Fourier transform (STFT) on a Hamming time-window. With this technique, the error bar in frequency is large (due to the small number of oscillations in the nonlinear regime around the saturation), namely of the order of 10-20\% (see Fig.~\ref{fig:NL-freq}-b). With such a big error-bar, no clear upward chirping is observed. Near the time of the saturation, i.e. $t\simeq 2.2 \, \Omega_i^{-1}$, only one mode is observed.
This is the condition of application of the direct technique of measurement of the frequency described above, where the frequency can be measured as the inverse of the period among peaks.

\begin{figure}[t!]
\begin{center}
\includegraphics[width=0.46\textwidth]{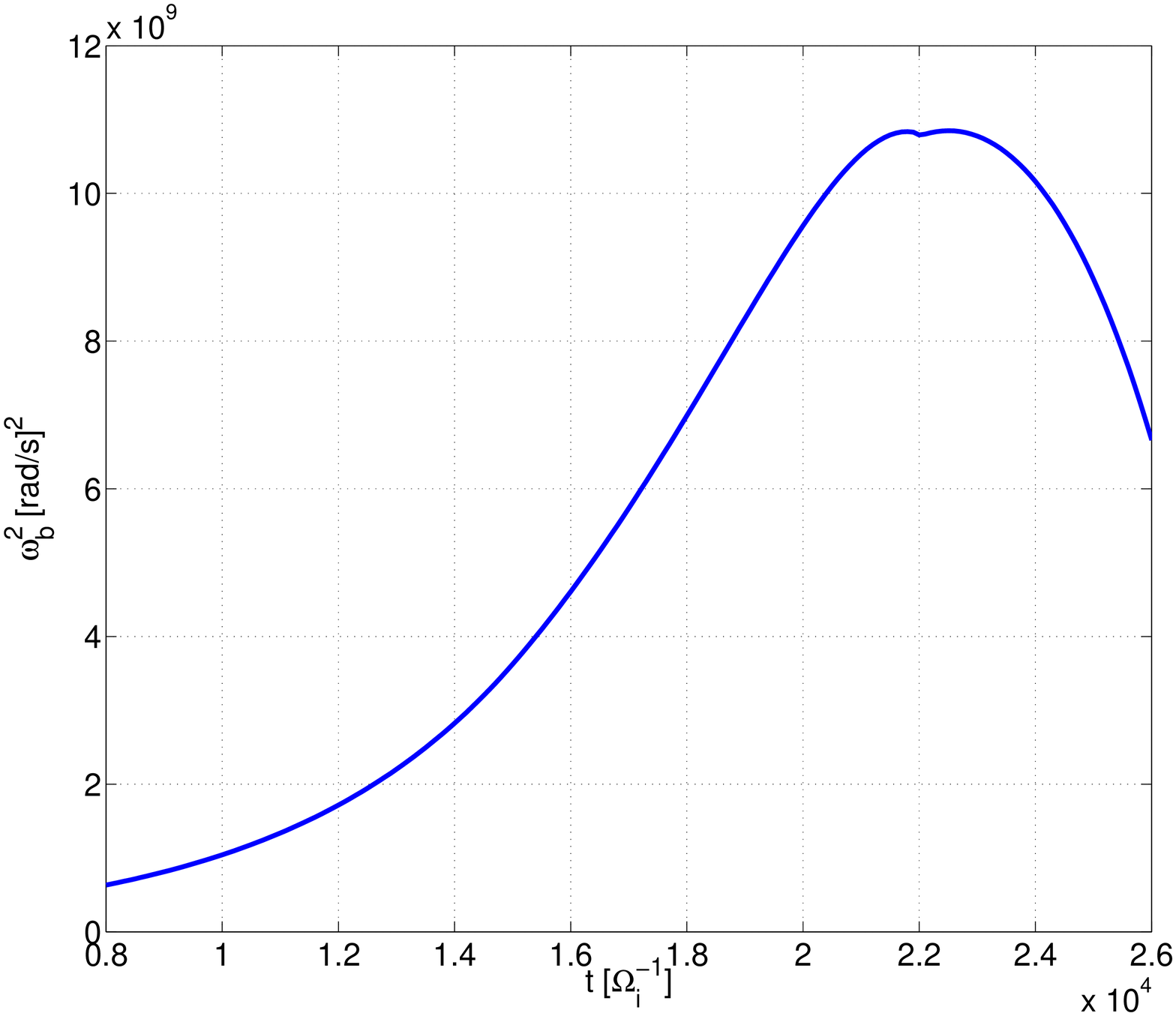}
\includegraphics[width=0.48\textwidth]{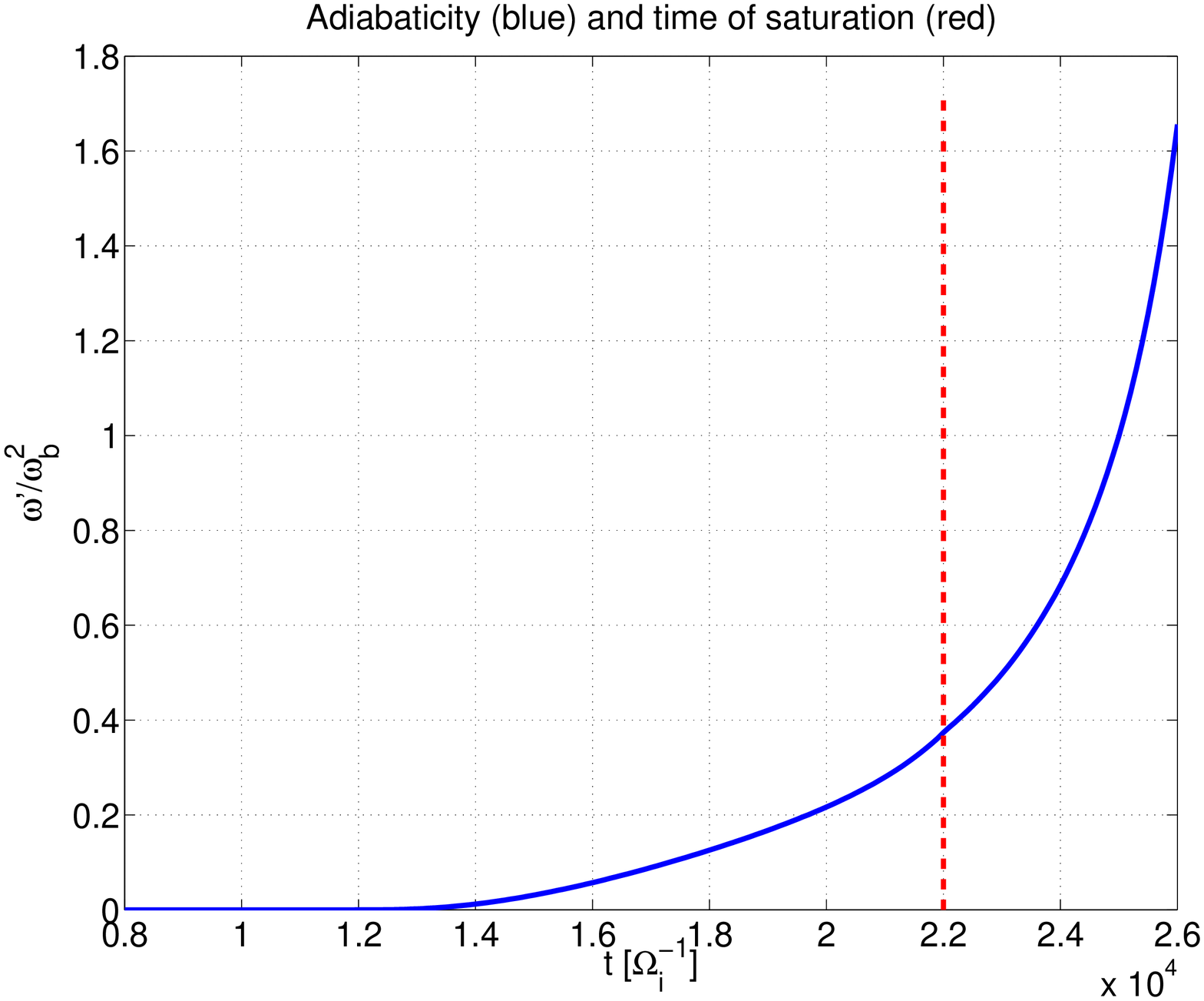}
\caption{Squared bounce frequency (left) and adiabaticity (right) for the EGAM with  $\rho^*=0.0078$, $n_{EP}/n_i=0.12$.\label{fig:adiabaticity}}
\end{center}
\end{figure}

As mentioned above, the EP bounce frequency $\omega_b$ depends on the mode amplitude. Its value for the considered simulation is shown in Fig.~\ref{fig:adiabaticity}-a. When the EGAM frequency  evolves slowly in time with respect to the inverse of the bounce frequency, then the EP can bounce back and forth many times, and this is called adiabatic dynamics.
The adiabaticity parameter, defined as $\omega'/\omega_b^2$, measures the level of adiabaticity of the dynamics. The time evolution of the adiabaticiy parameter for the considered simulation is depicted in Fig.~\ref{fig:adiabaticity}-b (with $\omega$ being the instantaneous mode frequency). 
A transition from adiabatic to non-adiabatic dynamics near the saturation is observed. In particular, near the saturation, the EP do not have the time to perform many bounces during the nonlinear modification of the wave. From this respect, the EGAM dynamics and the BPI are not in analogy. The details of the EGAM saturation mechanism will be investigated with a power-balance diagnostics in phase space, and discussed in a separate paper.

\section{Summary and discussion}
\label{sec:conclusions}

The importance of understanding the nonlinear dynamics of EGAM, i.e. energetic particle (EP) driven geodesic acoustic modes (GAM), is due to their possible role in interacting with turbulence and with the EP population present in tokamak plasmas as a result of nuclear reactions and/or heating mechanisms. In particular,  the level of the nonlinear saturation of EGAM is directly related to their efficiency in regulating turbulence or redistributing EP in phase space.

In this paper, we have investigated the problem of the nonlinear saturation of EGAM with the gyrokinetic particle-in-cell code ORB5, focusing on the wave-particle nonlinearity. Electrostatic collisionless simulations have been considered, with circular flux surfaces magnetic equilibria, and neglecting the kinetic effects of electrons.

The level of the saturated electric field has been shown to scale quadratically with the linear growth rate, similarly to the beam-plasma instability (BPI) in 1D uniform plasmas, and to Alfv\'en Eigenmodes (AE) near marginal stability.
We note that, in the case of beta-induced AE, due to the finite radial mode structure compared to resonant particle radial excursion, deviation from the $\delta E_r\propto \gamma_L^2$ is observed due to the competition of resonance detuning and radial decoupling~\cite{Wang12,Zhang12}.
A similar deviation is anticipated as one further increases the drive of EPs, from the correspondence of   EP pitch angle scattering by frequency chirping EGAMs and the radial wave-particle pumping by frequency chirping finite-n BAEs. This will be the content of our next publication.

We have also investigated the relationship of the bounce frequency of the EP in the field of the EGAM, with the linear growth rate, finding a linear dependence, similarly to the BPI. The linear coefficient, in the case of the EGAM problem, is proportional to the square root of the EGAM frequency, and is not a constant like in the beam-plamsa instability.
In the limit of $\omega_L/\omega_{GAM}\rightarrow 1$, which corresponds to the BPI problem when the mode frequency does not move sensibly from the value of the plasma frequency $\omega_p$, a unique value of the linear coefficient can be estimated: $\beta_0 = 2.66$, to be compared with the factor $\beta_{BPI}\simeq 3.2$ of the BPI.
This value of $\beta_0 = 2.66$ defines the EGAM in the regime of interest. The investigation of the dependence of $\beta$ on equilibrium parameters like the safety factor $q$, the magnetic flux surface elongation $e$, the EP distribution function, and the effect of kinetic electrons is left to  a future work.

Finally, we have investigated the temporal variation of the EGAM frequency during the saturation phase. The initial adiabatic regime, defined as the regime where the time derivative of the EGAM frequency is much smaller than the squared EP bounce frequency, has been shown to have a transition to a non-adiabatic regime at the saturation.
The exact analytical model of this phenomenon is still a work in progress and will be the first expansion of this work. The simplified picture is that the mode will chirp to maintain maximal exchange of power with the energetic particles~\cite{Chen16RMP}, while balancing the Landau damping. The mode frequency is at the same time bound by the wave dispersion relation and plasma equilibrium, making their simultaneous examination with the power exchange equation necessary for determining the rate and direction of chirping. The non-adiabatic behavior is enabled by the non-perturbative response of the energetic particles, which consists of significant distortion of the distribution function in the resonant region, as well as of the particle trapping and detrapping~\cite{Chen16RMP}. Differently, the adiabatic chirping is connected with either external driven fluctuations, or slow energetic particle redistribution, provided that the drive is sufficiently weak~\cite{Breizman97}.

\section*{Acknowledgements}
Interesting discussions with N. Carlevaro, G. Montani, F. Zonca on the nonlinear dynamics of the energetic particles are acknowledged. Interesting discussions with A. Mishchenko and B. Finney McMillan on the treatment of EGAMs with ORB5 are also acknowledged.
Part of this work has been carried out within the framework of the EUROfusion Consortium and has received funding from the Euratom research and training programme 2014-2018 under grant agreement No 633053, within the framework of the {\emph{Nonlinear energetic particle dynamics}} (NLED) European Enabling Research Project, WP 15-ER-01/ENEA-03. The views and opinions expressed herein do not necessarily reflect those of the European Commission.
Part of this work has been funded by the Centre de Coop\'eration Universitaire Franco-Bavarois - Bayerisch-Franz\"osisches Hochschulzentrum, for the grant FK 30\_15.
Simulations were performed on the Marconi supercomputer within the framework of the OrbZONE and ORBFAST projects. Part of this work was done while two of the authors (A. Biancalani and I. Novikau) were visiting LPP-Palaiseau, whose team is acknowledged for the hospitality. One of the authors (I.C.) wants to thank the Max Planck Princeton Center for the support during this work.

\end{document}